\documentclass[aps,pra,showpacs,amsmath,amsfonts,amssymb,superscriptaddress,preprint]{revtex4-1}

\usepackage[pdftex]{graphicx}
\usepackage{graphics}
\usepackage[latin1]{inputenc}
\usepackage{cleveref}
\usepackage{color}
\everymath{\displaystyle}

\newcommand{\affA}{Aix-Marseille University, Campus de Luminy, case 907,
CNRS Centre de Physique Théorique, UMR 7332, 13288 Marseille Cedex 09, France}

\begin{document}

\title{Semi-classical statistical approach to Fröhlich condensation theory}

\author{Jordane Preto}
\email{preto@cpt.univ-mrs.fr}
\affiliation{\affA}

\begin{abstract}
Fröhlich model equations describing phonon condensation in open systems of biological relevance are here reinvestigated
in a semi-classical non-equilibrium statistical context (with ``semi-classical'' it is meant
that the evolution of the system is described by means of classical equations with the addition of energy quantization). In particular, the assumptions 
that are necessary to deduce
Fröhlich rate equations are highlighted and we show how these hypotheses led us to write an appropriate form for the master equation.
As a comparison with known previous results, analytical relations with the Wu-Austin quantum Hamiltonian description are emphasized. 
Finally, we show how solutions of the master equation can be implemented numerically and outline some representative results of
the condensation effect. Our approach thus provides more information with respect to the existing ones, in what we are concerned with the time evolution of
the probability density functions instead of following average quantities.
\end{abstract}
\date{\today}


\maketitle

\section{Introduction}

Some decades ago, the study of open systems far from thermodynamic equilibrium showed, under suitable conditions, the emergence of self-organization.
Striking similarities were observed among very different physical systems having in common the fact of being composed of many non-linearly
interacting subsystems. When a control parameter, typically the energy input rate, exceeds a critical value then the subsystems act cooperatively
to self-organize in what is commonly referred to as a non-equilibrium phase transition. This is at variance with equilibrium phase transitions,
for which the transition from disorder to order is driven by a lowering of temperature -- the control parameter --below a critical value. Paradigmatic
examples of non-equilibrium phase transitions are provided by the laser transition and by the Rayleigh-Bénard convective instability, a quantum and a classical
system, respectively. In the early '80s of the last century, the emergence of collective properties in open systems was studied in a general and interdisciplinary
framework referred to as ``Synergetics'' \cite{haken}. This fascinating topic was pioneered in the late '60s by H. Fröhlich \cite{frohlich68}.

Fröhlich considered a system consisting in $z$ normal modes of frequency $\omega_k$ 
($\omega_1 < \omega_2 < ... < \omega_z$), each characterized by a discrete energy spectrum (phonons). Due 
to interactions with a surrounding heat bath, it was supposed that the dynamics of phonons
in each mode was governed by several kinds of stochastic events : $(i)$ linear events resulting in the absorption
or the emission of a single phonon from a particular normal mode, $(ii)$ non-linear events resulting in the simultaneous 
absorption of one phonon from a normal mode, \textit{and} emission
of one phonon to another mode. Besides interactions with the heat bath, each normal mode is assumed to be
fed energy by some external source. Denoting by $\langle N_k \rangle$ the \textit{average} number of 
phonons in the mode of frequency $\omega_k$, Fröhlich suggested that
the dynamics of the system was described by the following rate equations 

\begin{equation}\label{FRE}
\begin{array}{l}
\frac{d \langle N_k \rangle}{dt} = s_k + \varphi_k(\beta) \left(  \langle N_k \rangle +1  - 
\langle N_k \rangle e^{\beta \hbar \omega_k}  \right) + \vspace{0.3cm} \\
\hspace{2cm} \sum \limits_{j\neq k} \Lambda_{kj}(\beta) \left[\langle N_k + 1 \rangle \langle N_j \rangle 
- \langle N_k \rangle \langle N_j + 1 \rangle e^{\beta \hbar (\omega_k - \omega_j)} \right], \ \ k =1,...,z \ ,
\end{array}
\end{equation}

where $\beta = 1/kT$ with $T$ the temperature of the heat bath. Here, the first r.h.s. $s_k$
denotes the rate of external energy supply to the mode $k$ whereas the second and the third terms
account for the rate of change due to linear $(i)$ and non-linear $(ii)$ events, respectively (with  
appropriate temperature-dependent coupling constants $\varphi_k$ and $\Lambda_{kj}$, respectively).
Fröhlich solved analytically Eq. \eqref{FRE} in the stationary case
$\frac{d \langle N_k \rangle}{dt} = 0$, showing that the total 
number of phonons increased linearly by increasing the total rate of external energy supply
$\textstyle{\sum_k s_k}$ beyond a certain threshold, while 
non-linear events $(ii)$ tend to redistribute the energy excess
into the mode of lowest frequency. In other words, for sufficiently large value of the $s_k$'s, the
stationary state is shown to satisfy
$$
\langle N_1 \rangle \sim  \langle N \rangle
$$

where $\textstyle{\langle N \rangle= \sum\limits_{j=1}^{z} \langle N_j \rangle}$, \textit{i.e.}, it can be considered that 
all the components of the system oscillate in a collective way at the lowest frequency of the spectrum. This state of average excitation is better known as
\textit{Fröhlich condensation}. 

Since Fröhlich's first proposal \cite{frohlich68}, Fröhlich condensation has been investigated by many authors \cite{cifra}. 
In particular, a generic model of Hamiltonian was provided 
by Wu and Austin to derive Fröhlich rate equations [Eqs. \eqref{FRE}] from a microscopic quantum basis \cite{wu,pokornybook}. 
This type of Hamiltonian was then found to induce excitations propagating coherently in a similar way as Davydov's solitons \cite{tuszynski92, mesquita}.
From a practical point of view, Fröhlich condensation has been suggested to play a central role in the self-organization of biological polar structures
(here the set of normal modes arises from polar oscillations). This is the case, for instance, of microtubules and cell membranes that fulfilled the main requirements
of Fröhlich systems \cite{cifra, tuszynski11} (there the hydrolysis of adenosine triphosphate (ATP) or guanosine triphosphate (GTP) represents a significant source of 
external energy supply). Recently, Pokorný experimentally observed a strong excitation in the spectrum of vibration of microtubules localized in the $10$ MHz
range \cite{pokorny}. Among other applications of condensation, Fröhlich suggested that the excitation of a particular mode of polar oscillations in macromolecular systems
could lead to strong long-range dipole-dipole interactions. Applied to biological systems, it was expected that such forces would have
a profound influence on the displacement of \textit{specific} biological entities \cite{frohlich80, letter}, and thus on the initiation of a particular 
cascade of chemical events. While long-range interactions have been reported at the cellular level \cite{rowlands}, their possible role at the
biomolecular level still remains an open question \cite{article1}.

Now, coming back to Fröhlich theory, it should be stressed that most studies performed on the condensation -- including the derivation of Eqs. \eqref{FRE} -- are
based on a quantum Hamiltonian description akin to the one proposed originally by Wu and Austin. Given the semi-classical nature of the rate
equations postulated by Fröhlich, one can legitimately wonder how necessary a quantum description is to account for Fröhlich condensation
or even what are the ingredients needed to contemplate such a phenomenon. In the present paper, we propose to tackle this issue from a semi-classical
point of view (here ``semi-classical'' means
that the evolution of the system is described by means of classical equations with the addition of energy quantization). In section \ref{derivation}, Fröhlich rate equations \eqref{FRE} are derived from clear semi-classical arguments (\ref{OVPSS} and \ref{MVPSS}) 
and we show, in qualitative terms, how Fröhlich condensation can be ``deduced'' from these arguments (\ref{qualitative}). In section III, the classical
master equation which is at the grounds of the rate equations, is given from the main results found in section \ref{derivation}.
Then, the same equation is found to arise from the Wu-Austin Hamiltonian description under appropriate decoherence assumptions. Finally, as another original approach, 
Fröhlich condensation is statistically studied by estimating the solution of the master equation numerically [section IV]. A lot of informations
is thus provided that could, in our opinion, be of particular relevance to identify the phenomenon experimentally.





\section{Derivation of the rate equations}\label{derivation}

For the sake of readability, the derivation of Fröhlich rate equations has been itself splitted into two subsections. The first one (A)
deals specifically with the derivation of the second r.h.s. of Eq. \eqref{FRE}, \textit{i.e.} that part which accounts for 
linear events, whereas the second one (B) is about the derivation of the equations in their entirety. In the former case, it is enough to consider 
the presence of a single normal mode in the system. Thus, this will provide a good introduction to the second subsection
where the dynamics of all normal modes must be taken into account simultaneously in order to describe non-linear events.
Finally, a subsection (C) has been added as a complement to show how Fröhlich condensation can be ``predicted'' qualitatively on
the basis of the transition probabilities given in (A) and (B) (see below).

\subsection{One variable process}\label{OVPSS}

Although that point is usually not explicitly mentioned in the literature, Fröhlich equations are primarily based on the assumption 
that the time dependence of the number of phonons in each normal mode can be described as a homogeneous Markov process.
Here one normal mode of frequency $\omega_k$ is considered and we call $N_k(t)$ the stochastic process that accounts for the number of phonons 
in this mode. Under homogeneous Markov assumption it is possible to work with time-translation invariant
conditional probabilities $p(n_k,t |n'_k)$ such that

$$
p(n_k,t |n'_k)\equiv \mathrm{Prob}\{N_k(t+t') = n_k|N_k(t')=n'_k,\forall t'\ge 0 \},
$$

satisfies the following master equation \cite{gardiner}

\begin{equation}\label{OVME}
\partial_t p(n_k,t|n'_k) =  \sum \limits_{m_k} W(n_k|m_k)p(m_k,t|n'_k) - W(m_k|n_k)p(n_k,t|n'_k).
\end{equation}

Here, the transition probabilities

$$
W(n_k|m_k) = \lim_{\Delta t \rightarrow 0} \frac{p(n_k,\Delta t|m_k)}{\Delta t},
$$

are supposed to be well defined. Moreover, since Eq. \eqref{OVME} holds true irrespectively of the
initial condition $n'_k$, conditional probabilities $p(...,t|n'_k)$ will be substituted with standard ones $p(...,t)$ in what follows.

As mentioned in the heading of the section, only interactions that lead to the absorption or the emission of one phonon at a time
are considered here; thus $W(n_k|m_k)$ is zero except when $m_k =n_k \pm 1$ or $m_k=n_k$
(\textit{birth and death} process). Then, the equation for the evolution of the average number of phonons in the mode $k$ is easily deduced

\begin{equation}
\begin{array}{cl}
\frac{d \langle N_k \rangle}{dt} &=  \sum \limits_{n_k > 0}  n_k \partial_t p(n_k,t) \vspace{3mm} \\
& =  \sum \limits_{n_k > 0} n_k \left[ W(n_k|n_k-1) p(n_k-1,t) + W(n_k|n_k+1) p(n_k+1,t) \right. \\
& ~~~~~~~~~~~~~~~~~~~~~~~~~~~~ \left. - (W(n_k+1|n_k) + W(n_k-1|n_k)) p(n_k,t) \right].\\ \\
\end{array}
\end{equation}

With some appropriate substitutions in the sum index with suitable relabeling, one gets

\begin{equation}\label{OVRE}
\frac{d\langle N_k \rangle}{dt} = \sum \limits_{n_k > 0} 
\left[ W(n_k+1|n_k)- W(n_k-1|n_k) \right] p(n_k,t),
\end{equation}

where the boundary conditions

\begin{equation}\label{badcond}
 W(-1|0) = 0 , \ \mathrm{and} \ \ p(n_k,t) = 0, \ \mathrm{if} \ \  n_k<0,
\end{equation}

have been used. 

Now, let us look for an appropriate expression for $W(n_k+1|n_k)$ and $W(n_k-1|n_k)$ in terms of $n_k$. \textit{First}, we suppose that all
phonons are ``independent'' so that

\begin{equation}\label{perte}
W(n_k-1|n_k) = \alpha_k(\beta)  n_k,
\end{equation}

where $\alpha_k(\beta)$ is the probability per time unit (which is temperature-dependent in general) that one phonon is emitted from the mode $k$. 
\textit{Second}, $W(n_k + 1|n_k)$ can be obtained by noticing that a \textit{birth and death} process following boundary
 conditions \eqref{badcond} has a single stationary solution $p^s(n_k)$ through Eq. \eqref{OVME} 
(see the Appendix), and that this solution satisfies the detailed balance condition

\begin{equation}\label{OVDB}
W(n_k +1 |n_k)p^{s}(n_k) = W(n_k|n_k +1) p^{s}(n_k +1), \ \forall \ n_k\ge 0.
\end{equation}

Moreover, due to the contact between the system and the surrounding heat bath, it is required that the stationary
solution is given by a Bose-Einstein distribution

\begin{equation}
 p^{s}(n_k) = e^{-\beta \hbar \omega_k n_k}/Z, \ \ \mathrm{where} \ 
Z = \sum \limits_{n_k\ge 0} e^{-\beta \hbar \omega_k n_k} 
\end{equation}

is the one-normal mode partition function. Eq. \eqref{OVDB} then leads to :

\begin{equation}\label{OVTP}
W(n_k+1|n_k) = W(n_k|n_k+1) e^{-\beta \hbar \omega_k} \stackrel{\eqref{perte}}{=} \alpha_k(\beta) (n_k +1) e^{-\beta \hbar \omega_k}.
\end{equation}

Substituting this last equation and Eq. \eqref{perte} into Eq. \eqref{OVRE}, we finally get the second r.h.s.
of Eq. \eqref{FRE} : 

\begin{equation}\label{FFE}
\frac{d \langle N_k \rangle}{dt} = \varphi_k(\beta) \left(  \langle N_k \rangle +1  - 
\langle N_k \rangle e^{\beta \hbar \omega_k}  \right),
\end{equation}

where we have let $\varphi_k(\beta) \equiv \alpha_k(\beta) e^{-\beta \hbar \omega_k}$.

\subsection{Many variables process}\label{MVPSS}

We now want to consider a possible influence of all other normal modes on the mode of frequency $\omega_k$.
In this way, let $\boldsymbol{N}(t) = (N_1(t),N_2(t),...,N_z(t))$ be the vector of all Markovian processes describing
the number of phonons in each normal mode, so that the conditional probabilities are noted as 
$p({\boldsymbol{n}},t|\boldsymbol{n'})$, with ${\boldsymbol{n}}$, $\boldsymbol{n'}\in \mathbb{N}^z$, $\boldsymbol{n} = (n_1, n_2, ..., n_z)$ and 
$\boldsymbol{n'} = (n_1', n_2', ..., n_z')$.

In anticipation of a large number of interactions between the normal modes and the heat bath, we call $R^\mu, \ \mu= 1,...,s$,
all possible events that account for the simultaneous absorption(s) and/or emission(s) of phonon(s) from particular modes. 
Then, each event is characterized by a vector $\boldsymbol{r}^\mu$ whose element $r^\mu_{ i}$ 
($1\leq i \leq z$) is equal to the corresponding variation in the number of phonons in the mode of frequency $\omega_i$. Thus, the 
master equation takes the usual form \cite{gardiner}

\begin{equation}\label{MEME}
\partial_t p(\boldsymbol{n},t|\boldsymbol{n'}) =  \sum \limits_{\mu=1}^s W(\boldsymbol{n}|\boldsymbol{n} - \boldsymbol{r}^\mu)
p(\boldsymbol{n} - \boldsymbol{r}^\mu,t|\boldsymbol{n'}) - W(\boldsymbol{n}-\boldsymbol{r}^\mu|\boldsymbol{n})p(\boldsymbol{n},t|\boldsymbol{n'}).
\end{equation}

In this way, the evolution of the average number of phonons in the mode $k$ is immediately deduced 
(similarly to the previous subsection, initial conditions are omitted)

\begin{equation}\label{GRE}
\frac{d\langle N_k \rangle}{dt}  =  \sum \limits_{\mu=1}^s \sum \limits_{\boldsymbol{n}} n_k \left[ W(\boldsymbol{n}|\boldsymbol{n} - \boldsymbol{r}^\mu)
p(\boldsymbol{n} - \boldsymbol{r}^\mu,t) - W(\boldsymbol{n}-\boldsymbol{r}^\mu|\boldsymbol{n})p(\boldsymbol{n},t) \right],
\end{equation}

where the sum $\textstyle{\sum \limits_{\boldsymbol{n}}}$ corresponds to $\textstyle{\sum \limits_{n_1\ge 0}\sum \limits_{n_2\ge 0}...\sum \limits_{n_z\ge 0}}$. 
Besides, since events which do not result in a variation in the number of 
phonons in the mode $k$ have zero contribution to the equation, Eq. \eqref{GRE} can be readily simplified.

\begin{enumerate}
\item \textit{Second r.h.s. of Eq. \eqref{FRE}} : as mentioned above, each event $R^\mu$ corresponds to
the absorption or the emission of one phonon from a particular mode. Thus, there are only two events involving
the mode $k$ ($2z$ events overall) and they are given by $r^{\mu}_j = \pm \delta_{jk}$, for $j = 1,...,z$. This case corresponds to the case tackled
in the previous subsection whereby equation \eqref{FFE} was finally found.

\item \textit{Third r.h.s. of Eq. \eqref{FRE}} : here each event $R^\mu$ corresponds to the simultaneous 
absorption of one phonon from a normal mode, and emission
of one phonon from another mode. In particular, with regard to the mode $k$, there are $2(z-1)$ events ($2z(z-1)$ events overall), 
each characterized by $\boldsymbol{r}^{\mu}$ such that
\begin{equation}\label{events2}
\begin{array}{l}
\hspace{2.8cm} k \hspace{1.9cm} j \\
\hspace{2.8cm} \downarrow \hspace{1.9cm} \downarrow\\
\boldsymbol{r}^{\mu} = (0,....,0,\pm 1,0,....,0,\mp 1,0,....,0), \ \forall \ j\neq k, \
j \in \left[\kern-0.15em\left[1,z\right]\kern-0.15em\right],
\end{array}
\end{equation}

where $j \in \left[\kern-0.15em\left[1,z\right]\kern-0.15em\right]$ means $j=1, 2,...,z$.
For the sake of clarity in what follows, we shall note $W_{\boldsymbol{n}}(n_k;n_j|$ $n_k \pm 1;n_j \mp 1)$ 
the transition probabilities $W(\boldsymbol{n}|\boldsymbol{n}- \boldsymbol{r}^{\mu})$ associated with these events. Likewise,
$p(\boldsymbol{n} - \boldsymbol{r}^{\mu},t)$  will be written as
$p_{\boldsymbol{n}}(n_k \pm 1;n_j \mp 1,t)$ (in this way, $p(\boldsymbol{n},t)$ may be sometimes noted as $p_{\boldsymbol{n}}(n_k;n_j,t)$). 
In this way, Eq. \eqref{GRE} becomes

\begin{equation}
\begin{array}{rl}
\frac{d \langle N_k \rangle }{d t}  = &  \sum_{j\neq k}\sum \limits_{{\boldsymbol{n}}} n_k \left[
 W_{\boldsymbol{n}}(n_k;n_j|n_k - 1;n_j+1)p_{\boldsymbol{n}}(n_k - 1; n_j+1,t) \ +  \right.\\ 
& \hspace{4cm}  W_{\boldsymbol{n}}(n_k;n_j|n_k + 1;n_j-1) p_{\boldsymbol{n}}(n_k + 1; n_j-1,t) \ - \\ \\
&\hspace{4cm}  W_{\boldsymbol{n}}(n_k + 1;n_j - 1|n_k ;n_j) p_{\boldsymbol{n}}(n_k;n_j,t) \ -  \\ \\
& \left. \hspace{4cm}  W_{\boldsymbol{n}}(n_k - 1;n_j + 1|n_k ;n_j) p_{\boldsymbol{n}}(n_k;n_j,t) \right]. 
\end{array}
\end{equation}

Again, some appropriate substitutions on the sum indexes with suitable relabeling lead to

\begin{equation}\label{MVRE}
\begin{array}{l}
\frac{d\langle N_k \rangle}{dt} =  \sum_{j\neq k} \sum \limits_{{\boldsymbol{n}}}
\left[ W_{\boldsymbol{n}}(n_k+1;n_j-1|n_k;n_j) - \right. \\
\hspace{5cm} \left. W_{\boldsymbol{n}}(n_k-1;n_j+1|n_k;n_j) \right] p_{\boldsymbol{n}}(n_k;n_j,t),
\end{array}
\end{equation}

where the boundary conditions
\begin{equation}\label{badcond2}
\begin{array}{c}
W_{\boldsymbol{n}}(-1;n_j+1|0;n_j) = 0, \ W_{\boldsymbol{n}}(n_k+1;-1|n_k;0) = 0 , \ \mathrm{and} \ \\ \\
 \hspace{5cm}  \ p_{\boldsymbol{n}}(n_k;n_j,t) = 0, \mathrm{if} \ n_k \ \mathrm{or} \ n_j< 0,
\end{array}
\end{equation}

have been used.

As previously, we look for the expression of the transition probabilities. On each
normal mode $j\neq k$, we suppose that the phonons have the same probability $\gamma_{kj}(\beta,n_k)$ of being emitted 
while a phonon is absorbed in the mode $k$ (in general terms, this quantity depends on the temperature as well as
on the current number $n_k$ of phonons in the mode $k$). We thus get
\begin{equation}\label{gain-loss}
\begin{array}{c}
W_{\boldsymbol{n}}(n_k+1;n_j-1|n_k;n_j) \equiv W_{\boldsymbol{n}}(n_j-1;n_k+1|n_j;n_k) =\gamma_{kj}(\beta,n_k)  n_j, \\ \\
\hspace{7cm} \ \forall \ n_k, n_j\ge 0 \ 
\ \mathrm{and} \ \forall \ j\neq k, \
j \in \left[\kern-0.15em\left[1,z\right]\kern-0.15em\right].
\end{array}
\end{equation}

At this stage, it is worth mentioning that the transition probabilities considered thus far have been given in such a way
that the graph of the master equation is connected. Therefore, according to general results on master equations \cite{gardiner}, it can be found
that a stationary solution of Eq \eqref{MEME} exists and is unique. Moreover, in so far as no other external source is present, 
a stationary solution for which detailed balanced condition is fulfilled, always exists. In our case, this solution is thus unique.
Besides, since the system of normal modes interacts with the heat bath only, the stationary solution has to be given by a Bose-Einstein 
distribution. Thus

\begin{equation}\label{DB2}
\begin{array}{c}
W_{\boldsymbol{n}}(n_k+1;n_j-1|n_k;n_j)p^{s}_{\boldsymbol{n}}(n_k;n_j) = 
W_{\boldsymbol{n}}(n_k;n_j |n_k+1;n_j-1) p^{s}_{\boldsymbol{n}}(n_k+1;n_j-1) , \\ \\
\hspace{7cm} \ \forall \ n_k, n_j\ge 0 \ \ \mathrm{and} \ \forall \ j\neq k, \
j \in \left[\kern-0.15em\left[1,z\right]\kern-0.15em\right],
\end{array}
\end{equation}

with $p^{s}_{\boldsymbol{n}}(n_k;n_j) = \prod \limits_{i} e^{-\beta\hbar\omega_i n_i}/Z, \ \ \mathrm{where} \ 
Z = \prod \limits_{i}\sum \limits_{n_i\ge 0} e^{-\beta\hbar\omega_i n_i}$.

Using Eq. \eqref{gain-loss}, Eq. \eqref{DB2} becomes 

$$
\frac{\gamma_{kj}(\beta,n_k)}{n_k+1} = \frac{\gamma_{jk}(\beta,n_j-1)}{n_j}e^{-\beta\hbar(\omega_k -\omega_j)},
$$

which must depend on $k$ and $j$ only (\textit{i.e.}, not on $n_k$ nor on $n_j$). We finally deduce the properties
\begin{equation}\label{MVTP}
  \gamma_{kj}(\beta,n_k) = \Lambda_{kj}(\beta)(n_k + 1),
\end{equation}

with
\begin{equation}\label{DB}
 \Lambda_{jk}(\beta) = \Lambda_{kj}(\beta) e^{\beta\hbar(\omega_k -\omega_j)}, \ \  \forall \ j\neq k, \
j \in \left[\kern-0.15em\left[1,z\right]\kern-0.15em\right].
\end{equation}

Using these equations and Eq. \eqref{gain-loss} in Eq. \eqref{MVRE}, we obtain

 \begin{equation}
\frac{d\langle N_k \rangle}{dt}=  \sum \limits_{j\neq k} \Lambda_{kj}(\beta) \left[\langle (N_k + 1) N_j \rangle 
- \langle N_k (N_j + 1) \rangle e^{\beta \hbar (\omega_k - \omega_j)} \right], 
\end{equation}

which is exactly the third r.h.s. of Fröhlich equations \eqref{FRE} in so far as correlations between the normal modes can be 
neglected, \textit{i.e.}
$\langle N_k (N_j + 1) \rangle \simeq  \langle N_k \rangle \langle N_j + 1 \rangle $, or more specifically
$p({\boldsymbol{n}},t) \simeq \prod \limits_{i} p(n_i,t)$.

\item \textit{First r.h.s. of Eq. \eqref{FRE} (source term)} : the constant rate of external energy supply may 
be worked out by considering the external source as a heat bath with infinite temperature, with which the system 
interacts linearly. Thus, the rate of change in the mode $k$ due to the source is given by an equation similar to \eqref{FFE}

$$
\frac{d \langle N_k \rangle}{dt} = \varphi_k^{s}(\beta_s) \left(  \langle N_k \rangle + 1  - 
\langle N_k \rangle e^{\beta_s \hbar \omega_k}  \right)
$$

where $\beta_s \rightarrow 0$, so that 
\begin{equation}
\frac{d \langle N_k \rangle}{dt} = s_k \equiv \varphi_k^{s}(\beta_s)
\end{equation}
Here, $\beta_s = 1 /k T_s$ with $T_s$ the temperature of the source, and 
$\varphi_k^{s}(\beta_s)= \alpha_k^s(\beta_s) e^{-\beta_s \hbar \omega_k}$ where 
$\alpha_k^s(\beta_s)$ is the probability per time unit that a phonon is emitted from the mode $k$ to the source. 
\end{enumerate}

Finally, by gluing together all the events listed above (points 1. 2. and 3.), we deduce 

\begin{equation}\label{FRE2}
\begin{array}{l}
\frac{d \langle N_k \rangle}{dt} = s_k + \varphi_k(\beta) \left(  \langle N_k \rangle +1  - 
\langle N_k \rangle e^{\beta \hbar \omega_k}  \right) + \vspace{0.3cm} \\
\hspace{2cm} \sum \limits_{j\neq k} \Lambda_{kj}(\beta) \left[\langle N_k + 1 \rangle \langle N_j \rangle 
- \langle N_k \rangle \langle N_j + 1 \rangle e^{\beta \hbar (\omega_k - \omega_j)} \right], \ \ k =1,...,z \vspace{0.3cm} \\
\end{array}
\end{equation}

which correspond exactly to Eqs. (1) with $\varphi_k(\beta) = \alpha_k(\beta) e^{-\beta \hbar \omega_k}$ and $
\Lambda_{jk}(\beta) = \Lambda_{kj}(\beta) e^{\beta\hbar(\omega_k -\omega_j)}$.
\subsection{Qualitative approach to Fröhlich condensation} \label{qualitative}

As mentioned in the Introduction, Fröhlich condensation state is a stationary state for the system described 
by Eqs. \eqref{FRE2} achieved when the rates of energy supply $s_k$ exceed some threshold value. In this situation, the energy of the system is 
primarily located in the mode of lowest frequency, \textit{i.e.}, it can be 
found that $\langle N_1 \rangle \sim \langle N \rangle \equiv \sum_j \langle N_j\rangle$. 

Although the existence of this state has been emphasized both analytically \cite{frohlich68, frohlich80} and numerically \cite{pokornybook} from Eqs. \eqref{FRE2}, 
Fröhlich condensation may be highlighted more directly on the basis of the transition probabilities computed in the previous subsection. 
In that regard, it should be stressed that Fröhlich condensation is obtained as a consequence of the energy redistribution due
to the influence of non-linear interactions between the system and the heat bath \cite{frohlich80}. Now, we found in the previous subsection that
the probability per time unit that one phonon is emitted from the mode $j$ while one phonon is simultaneously absorbed in the mode $k$
reads as (see Eqs. \eqref{gain-loss} and \eqref{MVTP})

\begin{equation}\label{NLTP1}
W_{\boldsymbol{n}}(n_k+1;n_j-1|n_k;n_j) = \Lambda_{kj}(\beta)(n_k + 1)n_j. 
\end{equation}

Conversely, the probability per time unit that one phonon is emitted from the mode $k$ while one phonon is absorbed in the mode $j$
is simply $W_{\boldsymbol{n}}(n_k-1;n_j+1|n_k;n_j) = \Lambda_{jk}(\beta)n_k(n_j + 1)$. Then, using Eq. \eqref{DB} $\Lambda_{jk}$, one gets

\begin{equation}\label{NLTP2}
W_{\boldsymbol{n}}(n_k-1;n_j+1|n_k;n_j) = \Lambda_{kj}e^{\beta\hbar(\omega_k -\omega_j)}n_k(n_j + 1). 
\end{equation}

In particular, assuming that $n_k, n_j$ $\gg 1$, it appears that the probability that one phonon
is absorbed in the mode of \textit{lower} frequency through non-linear events is always greater than the probability that one phonon
is emitted from that mode. Indeed, when $n_k, n_j$ $\gg 1$, one gets  $(n_k + 1)n_j \sim n_k(n_j + 1)$. Thus, the ratio between
the transition probabilities \eqref{NLTP1} and \eqref{NLTP2} can be approximated as $e^{\beta\hbar(\omega_k -\omega_j)}$. If $\omega_k<\omega_j$,
$e^{\beta\hbar(\omega_k -\omega_j)}<1$ so that the probability of a simultaneous absorption in the mode $k$
and emission from the mode $j$ will be the larger one at any time (Eq. \eqref{NLTP1}). On the contrary,  if $\omega_k>\omega_j$,
$e^{\beta\hbar(\omega_k -\omega_j)}>1$ and the larger probability will be that of a simultaneous absorption in the mode mode $j$
and emission from the mode $k$ (Eq. \eqref{NLTP2}). Here, we see how relevant the condition of detailed balance through Eq. \eqref{DB}
is in order to get Fröhlich condensation. In particular, a wrong choice of the constants $\Lambda_{ij}$ could lead to a stationary state characterized by 
the excitation of modes of higher frequency in the spectrum of the system.

Coming back to the number of phonons in each mode, the condition for large values of 
$n_k, n_j$ can be fulfilled in so
far as the rates of energy supply $s_k$ are large enough. More specifically, it was shown at the end of the last subsection, that the interactions
between the system and the external source could be depicted similarly to the linear interactions between the system and the heat bath
with a condition of infinite temperature. For the latter, let us recall from subsection \ref{OVPSS} that the probability that one phonon is emitted from the mode $k$ 
(through linear events) reads as

\begin{equation}\label{LTP1}
W_{\boldsymbol{n}}(n_k-1|n_k) = \alpha_{k}(\beta)n_k, 
\end{equation}

while the probability of absorption is given by

\begin{equation}\label{LTP2}
W_{\boldsymbol{n}}(n_k+1|n_k) = \alpha_{k}(\beta)(n_k+1)e^{-\beta \hbar \omega_k}. 
\end{equation}

Here, in the absence of source, the condition for large values for the $n_k$'s can never be fulfilled because
if the $n_k$'s become too large, emission will be favored against absorption. On the contrary, when only
the external source is considered, one has $\beta = \beta_s \rightarrow 0$ so 
that according to \eqref{LTP1} and \eqref{LTP2} (here $\alpha_{k}$ must be changed in $s_k$) absorption is always favored. Thus, in 
considering the presence of both heat bath and external source, the possibility of working with large number of phonons
in each normal mode will depend to some extent on whether the ratio $s_k/\alpha_k$ is large or not. In this way, supposing that $s_k \gg \alpha_k$
for all $k$, one will naturally get $n_k \gg 1$ after a certain time so that non-linear interactions will redistribute
the energy into the mode of lowest frequency.

\section[The original ME and the relation to the WAH description]{The original Fröhlich Master equation and the 
relation to the Wu-Austin Hamiltonian description}

In the previous section, we have been able to derive the Fröhlich rate equations \eqref{FRE2} on the 
basis of purely semi-classical arguments. To summarize, it was assumed that :

\begin{enumerate}
\item The number of phonons in each normal mode of the system (as a function of time) could be represented as a (homogeneous) Markov process. This
property allowed us to describe the evolution of the system on the basis of a classical master equation \eqref{MEME}.
\item Due to the discrete nature of the energy (phonons), each process could be described as a \textit{birth and death}
process.
\item On each normal mode, all phonons could be considered independent. In addition, according to the detailed
balance condition and the requirement of Bose-Einstein stationary distributions of phonons when only one of both the external source and the
heat bath is present, the transition probabilities that take place in the master equation have been worked out.
\end{enumerate}
In the end, the previous calculations allowed us to derive not only the Fröhlich rate equations
but also the original classical master equation from which the former can be deduced

\begin{equation}\label{FME}
\partial_t p({\boldsymbol{n}},t) = \partial_t p_s({\boldsymbol{n}},t) + \partial_t p_\alpha({\boldsymbol{n}},t) + \partial_t p_\Lambda({\boldsymbol{n}},t).
\end{equation}

Here, $p_s({\boldsymbol{n}},t)$, $p_\alpha({\boldsymbol{n}},t)$ and $p_\Lambda({\boldsymbol{n}},t)$ 
are the probability density functions of the number of phonons in all normal modes of the system
(let us emphasize again that $\boldsymbol{n} =(n_1,n_2,...,n_z)$) 
whose evolutions are respectively due to :
\begin{itemize}
\item (Linear) events resulting from interactions between the system and the external source;
\item Linear events $(i)$ resulting from interactions between the system and the heat bath;
\item Non-linear events $(ii)$ resulting from interactions between the system and the heat bath.
\end{itemize}
By substituting Eqs. \eqref{LTP1} and \eqref{LTP2}
into Eq. \eqref{MEME}, we get the master equation for $p_\alpha({\boldsymbol{n}},t)$

\begin{equation}\label{FME1}
\begin{array}{ll}
\partial_t p_\alpha({\boldsymbol{n}},t) & = \sum_i \alpha_i(\beta) \left\{ [(n_i +1) p_{\boldsymbol{n}}(n_i +1,t) - n_i p_{\boldsymbol{n}}(n_i,t) ] 
\ + \right.  \\
& \hspace{3cm} \left. + e^{- \beta \hbar \omega_i} [ n_i  p_{\boldsymbol{n}}(n_i - 1,t) - (n_i + 1) p_{\boldsymbol{n}}(n_i,t)] \right\}.
\end{array}
\end{equation}

The master equation for $p_s({\boldsymbol{n}},t)$ is found in a similar way by letting  $\beta = \beta_s \rightarrow 0$
and changing $\alpha_{i}$ in $s_i$ 
\begin{equation}\label{FMES}
\partial_t p_s({\boldsymbol{n}},t)  = \sum_i s_i(\beta_s) \left\{ (n_i +1) p_{\boldsymbol{n}}(n_i +1,t) + 
n_i  p_{\boldsymbol{n}}(n_i - 1,t) - (2 n_i + 1) p_{\boldsymbol{n}}(n_i,t)    \right\}.
\end{equation}

Let us recall from subsection \ref{MVPSS} that $p_{\boldsymbol{n}}(m_i,t)$ is the probability of getting 
$n_1$ phonons in the mode of frequency $\omega_1$, $n_2$ phonons in the mode of frequency $\omega_2$,..., $n_z$ phonons in the mode of frequency $\omega_z$ 
but $m_i$ phonons (instead of $n_i$ phonons) in the mode of frequency $\omega_i$. A similar notation $p_{\boldsymbol{n}}(m_i;m_j,t)$ is used when the 
exception involves two modes. Naturally, $p_{\boldsymbol{n}}(n_i,t)$ and $p_{\boldsymbol{n}}(n_i;n_j,t)$ correspond to $p({\boldsymbol{n}},t)$.

Finally, the master equation corresponding to non-linear events is given by substituting 
Eqs. \eqref{NLTP1} and \eqref{NLTP2} into Eq. \eqref{MEME} so that

\begin{equation}\label{FME2}
\begin{array}{ll}
\partial_t p_\Lambda({\boldsymbol{n}},t) & = \sum_{i,j>i} \Lambda_{ij}(\beta) n_i( n_j + 1)p_{\boldsymbol{n}}(n_i -1;n_j+1,t) \ - \\
& \hspace{3cm} \Lambda_{ij}(\beta) (n_i +1 ) n_j  p_{\boldsymbol{n}}(n_i;n_j,t) \ \vspace{0.3cm} + \\
& \hspace{3cm} \Lambda_{ji}(\beta) (n_i +1 ) n_j  p_{\boldsymbol{n}}(n_i +1;n_j-1,t) \ \vspace{0.3cm} - \\
& \hspace{4cm} \Lambda_{ji}(\beta) n_i (n_j +1) p_{\boldsymbol{n}}(n_i;n_j,t),
\end{array}
\end{equation}

with $\Lambda_{ji}(\beta) = \Lambda_{ij}(\beta) e^{\beta\hbar(\omega_i -\omega_j)}$.

Eq. \eqref{FME} together with Eqs. \eqref{FME1}, \eqref{FMES} and \eqref{FME2} are the
original classical master equations from which the Fröhlich rate equations \eqref{FRE2} may 
be deduced. In the Introduction, it was mentioned that Eqs \eqref{FRE} could be
derived from a quantum Hamiltonian that was originally introduced
by Wu and Austin \cite{wu,pokornybook}. More specifically, it can be found that the above master equations
are no more than an intermediate step of that derivation. In that regard, let us also emphasized that 
the possibility of deducing classical master equations (also called Pauli master equations) from
a microscopic Hamiltonian is well-known in the realm of quantum open systems \cite{louisell,zwanzig}. 
In the case of Fröhlich theory, that correspondence has not been given much attention in the literature.
From this point of view as well as to provide a consistency check of Eqs. \eqref{FME} to \eqref{FME2},
let us emphasize how the Fröhlich master equations can be deduced from the quantum approach
given by Wu and Austin. 

To that purpose, we recall that the Wu-Austin Hamiltonian \cite{wu} is given by

$$
H = H_0 + H_{int},
$$

where $H_0$ stands for the Hamiltonian of the system of normal modes, the heat bath plus the external source in the absence of interactions,
and has the following form :

$$
H_0=  \sum \limits_{i=1}^{z} \hbar \omega_i a_i^\dagger a_i + \sum \limits_{k=1}^{z_b} 
\hbar \Omega_k b_k^\dagger b_k + \sum \limits_{p=1}^{z_s} \hbar \Omega'_p c_p^\dagger c_p.
$$

Here, $a_i^\dagger$, $a_i$ are the creation and annihilation operators associated
with the normal mode of frequency $\omega_i$ of the system,  $b_k^\dagger$,$b_k$ are the creation and annihilation operators associated with 
the normal mode of frequency $\Omega_k$ of the heat bath, and $c_p^\dagger$, $c_p$ are the creation and annihilation operator associated with the
normal mode of frequency $\Omega'_p$ of the source. In parallel, $H_{int}$ is the interaction Hamiltonian such that

\begin{equation}\label{wu}
H_{int} =  \sum \limits_{i,k} \lambda_{i,k} a_i^\dagger b_k + \sum \limits_{i,j,k} \chi_{i,j,k} a_i a_j^\dagger  
b_p + \sum \limits_{i,p} \xi_{ip} a_i^\dagger c_p  + H.c.,
\end{equation}


Since one is interested in the statistical properties of 
observables related to the system only, it is convenient to work with the reduced density operator $S(t)$,
which is defined as the usual density operator traced over the states related to the heat bath and the source \cite{louisell}. 
Assuming that the dynamics of each normal mode verifies the Markov property, it may be shown \cite{louisell,hirsch} 
that second perturbation theory in the interaction picture applied to the interaction Hamiltonian \eqref{wu} actually yields

\begin{equation}\label{QME}
 \begin{array}{ll}
  \frac{dS}{dt} = & - \ \frac{1}{2} \sum_i \left \{ (\phi_i r_i + \sigma_i ) \left(a_i a_i^\dagger S + S a_i a_i^\dagger 
-2 a_i^\dagger S a_i \right) \ +  \right.\\
& \left . \hspace{2cm} [\phi_i (r_i + 1) + \sigma_i ] \left(a_i^\dagger a_i S + S a_i^\dagger a_i
-2 a_i S a_i^\dagger \right) \right\}  \vspace{0.3cm} \\ 
& - \frac{1}{2} \sum \limits_{\substack{i, j, i',j' \\ j>i, j'>i' \\ i-i'=j-j'}}  \frac{\pi}{\hbar^2}
\chi_{i,j,j-i} \chi^*_{i',j',j'-i'} \times \\
& \hspace{2cm}\left[ \left(a_i a_j^\dagger a_{j'} a_{i'}^\dagger S +
 S a_i a_j^\dagger a_{j'} a_{i'}^\dagger \ - 2 a_{j'}  a_{i'}^\dagger S a_i a_j^\dagger \right) 
(r_{j-i} + 1) \ + \right.  \vspace{0.3cm} \\ 
& \hspace{4cm}  \left. \left(a_{j'} a_{i'}^\dagger a_i a_j^\dagger  S + S a_{j'} a_{i'}^\dagger a_i a_j^\dagger 
- 2 a_i a_j^\dagger S a_{j'}  a_{i'}^\dagger \right) r_{j-i} \right],
 \end{array}
\end{equation}

where :

\begin{equation}\label{wu_notations}
 \begin{array}{l}
 \phi_i = \frac{\pi}{\hbar^2} \left( |\lambda_{i,i}|^2 + |\xi_{i,i}|^2  \right) \ , \ \ 
\sigma_i = \frac{\pi}{\hbar^2} |\xi_{i,i}|^2 \left( \langle c_i^\dagger c_i \rangle_s - \langle b_i^\dagger b_i \rangle_b \right).
 \end{array}
\end{equation}

Here, $\langle ... \rangle_s$ and $\langle ... \rangle_b$ are averages performed over states related to the source and the heat bath
respectively. 
Now, the probability of observing $n_1$ phonons in the first mode,  $n_2$ phonons in the second, and so on, corresponds to
the matrix element $\langle {\boldsymbol{n}} |S(t)| {\boldsymbol{n}} \rangle := p({\boldsymbol{n}},t)$, where 
 $|{\boldsymbol{n}}\rangle = | n_1 \rangle ...|n_z \rangle$. Naturally, $| n_i \rangle$ is the eigenvector 
of the operator $\hat{n}_i = a_i^\dagger a_i$. Considering only the first sum in Eq. \eqref{QME} (that stems from the linear terms in 
the Hamiltonian \eqref{wu}), one can easily deduce the equation :

\begin{equation}\label{PE_part1}
 \begin{array}{ll}
\partial_t p({\boldsymbol{n}},t) & =  - \sum_i (\phi_i r_i + \sigma_i )  \left[ (n_i + 1) p_{\boldsymbol{n}}(n_i,t) -
 n_i  p_{\boldsymbol{n}}(n_i - 1,t) \right]
\ +   \\
& \hspace{3cm}  + (\phi_i (r_i + 1) + \sigma_i )  \left[ n_i p_{\boldsymbol{n}}(n_i,t) - (n_i +1) p_{\boldsymbol{n}}(n_i +1,t)  \right]
 \end{array}
\end{equation}

In particular when no source is present, one simply gets 
$$
 \phi_i r_i + \sigma_i = \frac{\pi}{\hbar^2} |\lambda_{i,i}|^2 \langle b_i^\dagger b_i \rangle_b
= \frac{\pi}{\hbar^2}|\lambda_{i,i}|^2\frac{1}{e^{\beta \hbar \omega_i} -1} \  \ \ \text{and}  \ \ \
 \phi_i (r_i +1) + \sigma_i = \frac{\pi}{\hbar^2} |\lambda_{i,i}|^2 \frac{e^{\beta \hbar \omega_i}}{e^{\beta \hbar \omega_i} -1},
$$
as the heat bath is required to be at thermal equilibrium with a temperature $\beta = 1/kT$.
Using these two last expressions  into Eq. \eqref{PE_part1}, we obtain :

\begin{equation}
\begin{array}{ll}
\partial_t p({\boldsymbol{n}},t) & = \sum_i \frac{\pi}{\hbar^2} |\lambda_{i,i}|^2 
\frac{e^{\beta \hbar \omega_i}}{e^{\beta \hbar \omega_i} -1} 
\left\{ [(n_i +1) p_{\boldsymbol{n}}(n_i +1,t) - n_i p_{\boldsymbol{n}}(n_i,t) ] \ + \right.  \\
& \hspace{3cm} \left. + e^{- \beta \hbar \omega_i} [ n_i  p_{\boldsymbol{n}}(n_i - 1,t) - (n_i + 1) p_{\boldsymbol{n}}(n_i,t)] \right\}
\end{array}
\end{equation}

which is exactly the classical master equation \eqref{FME1} with $\alpha_i(\beta) =\frac{\pi}{\hbar^2} |\lambda_{i,i}|^2 
\frac{e^{\beta \hbar \omega_i}}{e^{\beta \hbar \omega_i} -1}$.
Naturally, the classical master equation \eqref{FMES} related to the source is found in a similar way by considering only the $\xi$-terms in Eq. \eqref{QME}  and
by using the condition $\langle c_i^\dagger c_i \rangle_s = \frac{1}{e^{\beta_s \hbar \omega_i} -1}$ with $\beta_s \hbar \omega_i \ll 1$.

Although no further assumption has been needed to get Eqs. \eqref{FME1} and \eqref{FMES} from Eq. \eqref{QME},
the master equation \eqref{FME2} related to non-linear interactions may only be deduced by first supposing that all quantum coherences
have been destroyed in the system, \textit{i.e} :

\begin{equation}\label{decoherence}
\langle {\boldsymbol{n}} |S(t) | \boldsymbol{m} \rangle = p({\boldsymbol{n}},t) \delta_{{\boldsymbol{n}},\boldsymbol{m}}, \ \ \ \forall \boldsymbol{n}, \boldsymbol{m}
\end{equation}

where $\delta_{{\boldsymbol{n}},\boldsymbol{m}}= \delta_{n_1,m_1}\delta_{n_2,m_2} ... \delta_{n_z,m_z}$. Surprisingly, 
although Eq. \eqref{decoherence} is a quite usual assumption when deriving classical master equations from a microscopic Hamiltonian, it was never 
clearly put in evidence in the case of Fröhlich theory.

Finally, one gets (we recall that according to Eq. \eqref{QME}, the condition $i-i'=j-j'$ is required) :

$$
\langle {\boldsymbol{n}} | a_i a_j^\dagger a_{j'} a_{i'}^\dagger S(t) | {\boldsymbol{n}} \rangle =
 \delta_{i,i'} \delta_{j,j'} (n_i + 1) n_j p_{\boldsymbol{n}}(n_i;n_j,t) = 
\langle {\boldsymbol{n}} |S(t) a_i a_j^\dagger a_{j'} a_{i'}^\dagger | {\boldsymbol{n}} \rangle,
$$

$$
\langle {\boldsymbol{n}} |a_{j'}  a_{i'}^\dagger S a_i a_j^\dagger| {\boldsymbol{n}} \rangle =  
 \delta_{i,i'} \delta_{j,j'} n_i (n_j + 1)  p_{\boldsymbol{n}}(n_i-1;n_j + 1,t),
$$

$$
\langle {\boldsymbol{n}} | a_{j'} a_{i'}^\dagger a_i a_j^\dagger  S(t) | {\boldsymbol{n}} \rangle =
 \delta_{i,i'} \delta_{j,j'} n_i (n_j + 1)  p_{\boldsymbol{n}}(n_i;n_j,t) = 
\langle {\boldsymbol{n}} |S(t) a_{j'} a_{i'}^\dagger a_i a_j^\dagger | {\boldsymbol{n}} \rangle ,
$$

and

$$
\langle {\boldsymbol{n}} | a_i a_j^\dagger S a_{j'}  a_{i'}^\dagger| {\boldsymbol{n}} \rangle =  
 \delta_{i,i'} \delta_{j,j'} (n_i + 1) n_j p_{\boldsymbol{n}}(n_i + 1;n_j - 1,t).
$$

From the second sum in Eq. \eqref{QME} (terms in $\chi$) and the relations $r_{j-i} = \frac{1}{e^{\beta\hbar(\omega_j - \omega_i)} -1}$ and
$r_{j-i} + 1 = \frac{e^{\beta\hbar(\omega_j - \omega_i)}}{e^{\beta\hbar(\omega_j - \omega_i)} -1}$, one obtains :

\begin{equation}
\begin{array}{ll}
\partial_t p({\boldsymbol{n}},t) & =  \sum_{i,j>i}  \frac{\pi}{\hbar^2} |\chi_{i,j,j-i}|^2 
\frac{1}{e^{\beta\hbar(\omega_i - \omega_j)} -1} \left\{ (n_i +1 ) n_j  p_{\boldsymbol{n}}(n_i;n_j,t) \ + \right. \\
& \hspace{3cm} n_i( n_j + 1)p_{\boldsymbol{n}}(n_i -1;n_j+1,t)   \ \vspace{0.3cm}  + \\
& \hspace{3cm} e^{\beta \hbar(\omega_i - \omega_j)} \left( (n_i +1) n_j  p_{\boldsymbol{n}}(n_i +1;n_j-1,t)  \right. \ \vspace{0.3cm} - \\
& \hspace{4cm} \left. \left. n_i (n_j +1) p_{\boldsymbol{n}}(n_i;n_j,t) \right) \right\},
\end{array}
\end{equation}
that is exactly the master equation \eqref{FME2} accounting for the non-linear coupling between the normal modes with 
$\Lambda_{ij}(\beta) =  \frac{\pi}{\hbar^2} |\chi_{i,j,j-i}|^2 
\frac{1}{e^{\beta\hbar(\omega_i - \omega_j)} -1}$ and $\Lambda_{ji}(\beta) =\frac{\pi}{\hbar^2} |\chi_{i,j,j-i}|^2 
\frac{e^{\beta \hbar(\omega_i - \omega_j)} }{e^{\beta\hbar(\omega_i - \omega_j)} -1}$. Therefore, we see that
the condition for detailed balance [Eq. \eqref{DB}] is adequately verified. As a final complement, let us mention that the Fröhlich master
equation can be deduced not only from the Hamiltonian \eqref{wu} but also from a large family of Hamiltonians for which the operators
describing the thermal bath and the source can be very complex objects as various combinations of bosonic and/or fermionic, creation
and/or annihilation operators \cite{turcu}. In the end, only the expressions of the parameters $s$, $\varphi$ and $\chi$ will differ from the ones
given in this section. We refer the reader
interested in a further detailed investigation on the quantum Hamiltonian formulation of Fröhlich theory to references \cite{wu,pokornybook,tuszynski92,mesquita} .

\section{Numerical computations}

It should be stressed that most theoretical studies on Fröhlich condensation are 
based on the numerical integration of the rate equations \eqref{FRE2} whereas the issue of the fluctuations of the number of phonons in each mode is
largely unexplored. From this point of view, a study of the distributions of the occupation numbers could be of utmost importance to identify Fröhlich 
condensation experimentally at least in qualitative terms.
As emphasized above, the time evolution of these distributions is provided by Eqs. \eqref{FME} to \eqref{FME2}.
These equations, being master equations with many variables, 
are extremely difficult to solve analytically as the number of involved processes (here the number of normal modes) increases. Nevertheless,
several algorithms allow to estimate the solution of such equations numerically. The most popular one, proposed by Gillespie,
consists in generating stochastic trajectories governed by the master equation on the basis of Monte-Carlo procedures \cite{gillespie}.
Eventually, these trajectories can be used to estimate the expected distribution at any time $t$ as well as the related moments.

In this context, we investigate numerically Fröhlich condensation through two different approaches. The \textit{first} one, as a consistency check, 
involves the integration of the rate equations \eqref{FRE2} so that the evolution of the average number of phonons of each normal mode is
followed for different values of the source parameters $s_k$. The \textit{second} approach consists in generating stochastic 
trajectories whose evolution is governed by the Fröhlich master equation [Eqs. \eqref{FME} to \eqref{FME2}], by means of the Gillespie algorithm. 
By collecting the results in the form of histograms, it is thus possible to estimate the distribution of phonons for any normal mode at any time $t$.

\begin{figure}
    \includegraphics[width=4.6in]{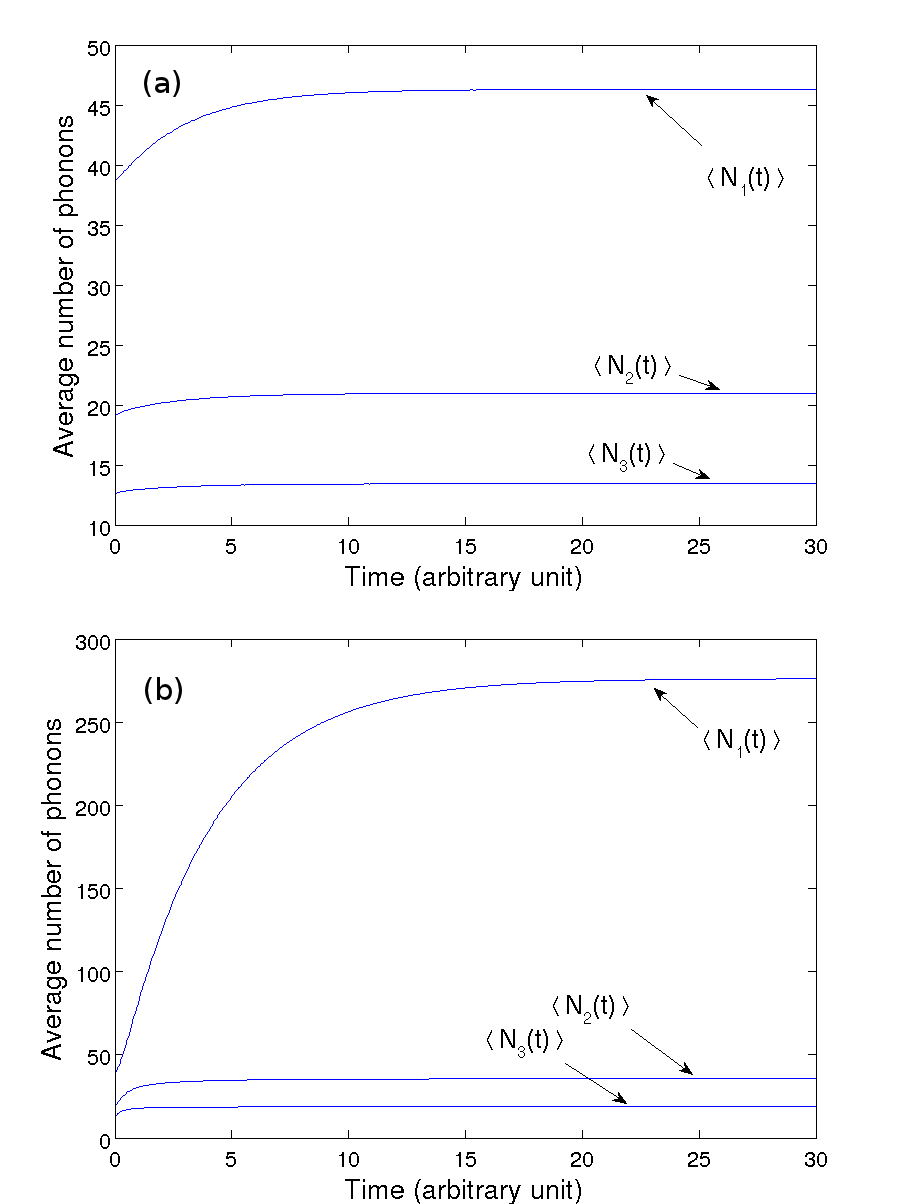}
\caption{(Color online) Time evolutions of the average number of phonons $\langle N_k (t) \rangle$ in the three modes of lowest frequency
of a system of four modes obtained by integrating Eqs. \eqref{FRE2}.
The frequencies of the normal modes are $\omega_i = i . 10^{12}$ Hz \cite{note2}, $i =1,...,4$  whereas the coupling constants associated with the 
interactions between the system and the heat bath are given by $\alpha_k = 10$ for all $k$ and $\Lambda_{kj} = 1$ for $k<j$ 
(let us recall that the relation $\Lambda_{jk} = \Lambda_{kj}e^{\beta \hbar (\omega_k - \omega_j)}$ is required). Initially, the system is supposed
to be at thermal equilibrium so that $\langle N_k (t=0) \rangle$ is given by Planck formula \eqref{planck}. The main change
between both figures is the rate of energy supply : (a) $s_k$ = 1, (b) $s_k$ = 20, for all $k$. All coupling constants
$\alpha$, $\Lambda$ and $s$ are given in (time arbitrary unit)$^{-1}$, as discussed in the text.}\label{FREfig}
 \end{figure}

Some representative results of Fröhlich condensation using the first approach are
reported in Figure \ref{FREfig}. The system is initially at thermal equilibrium, \textit{i.e}, the average number of phonons is given at $t=0$ by the Planck 
formula :

\begin{equation}\label{planck}
\langle N_k(t=0) \rangle = \frac{1}{e^{\beta \hbar \omega_k} - 1}, \ \ \forall k.
\end{equation}

Then, each solution tends to a stationary state with a convergence rate that is mainly depending on the values of 
the coupling constants $s_k$, $\alpha_k$, $\Lambda_{kj}$. Since we are interested in a 
qualitative description of the condensation, we are not committed to an exhaustive exploration of the parameter space
(in particular, the coupling constants are given in (time arbitrary units)$^{-1}$) \cite{note}.
Let us simply specify that we required $\Lambda_{kj} \ll \alpha_k$ $\forall k,j$ because events related to linear interactions are more likely to arise than events 
related to non-linear ones. Also, the coefficients $\Lambda_{kj}$ have been fixed according to the relation 
$\Lambda_{jk} = \Lambda_{kj}e^{\beta \hbar (\omega_k - \omega_j)}$ which follows a theoretical condition of detailed balance 
(see section \ref{MVPSS}). Now, we see from Figure \ref{FREfig} how Fröhlich condensation takes place in
average \text{terms :} a situation wherein  $s_k < \alpha_k$, for all $k$, leads to a stationary state relatively close to the initial (thermal equilibrium) condition 
(see Figure \ref{FREfig}(a)). Alternatively, a situation where $s_k \gg \alpha_k$ holds for the majority of normal modes leads to a stationary state characterized by a large 
number of phonons populating the mode of lowest frequency, \textit{i.e.}, $\langle N_1(\infty) \rangle \gg \langle N_k(\infty) \rangle $ for all $k>1$
(see Figure \ref{FREfig} (b)). Besides, the values of $\langle N_k(\infty) \rangle$ with $k>1$ still remain close to the thermal equilibrium values. 
 
\begin{figure}
    \includegraphics[width=4.6in]{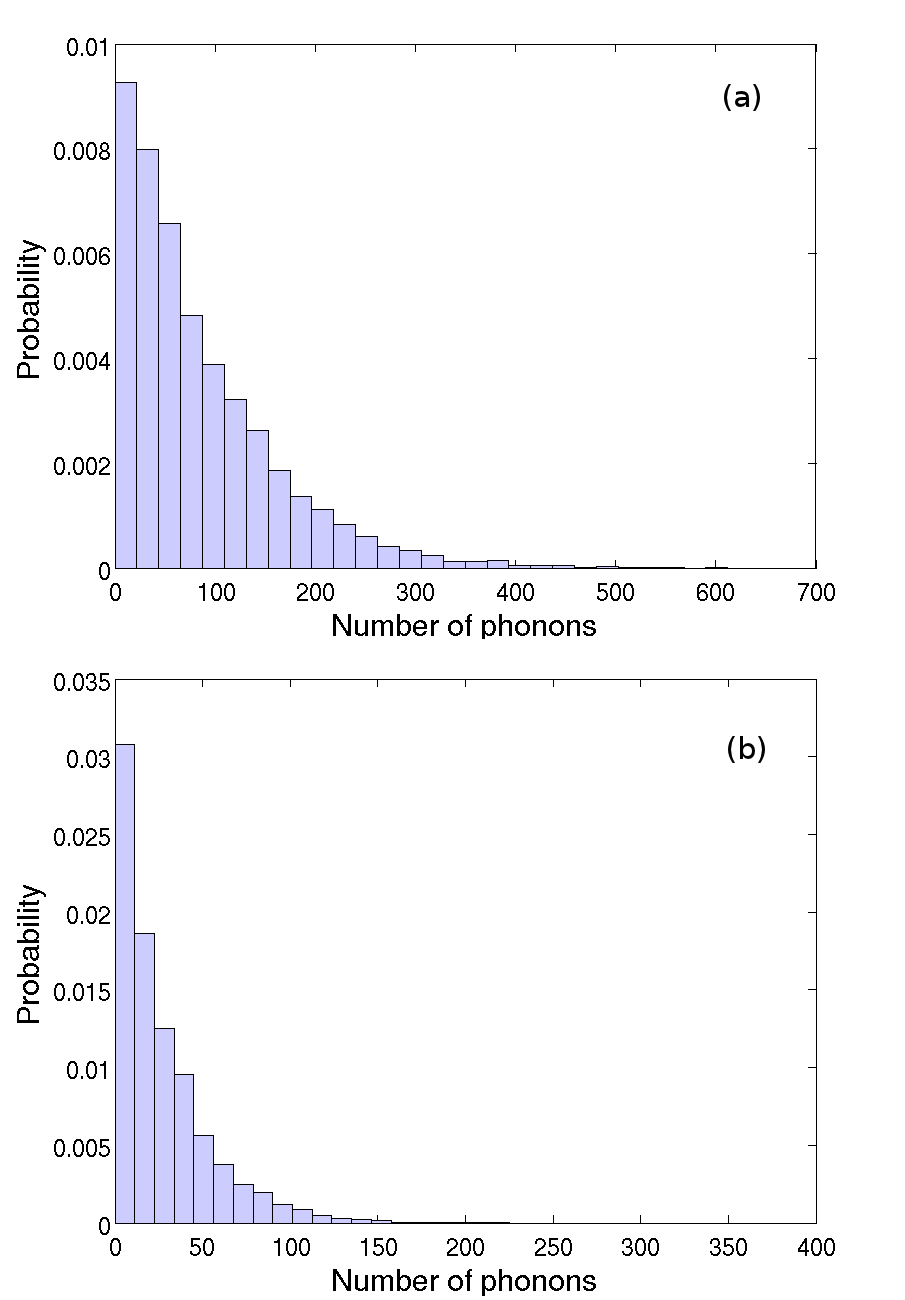}
\caption{(Color online) Normalized histograms of the number of phonons in the first two modes of a system of four modes at $t=30$ arbitrary units (= stationary state) :
(a) mode of frequency $\omega_1$ (b) mode of frequency $\omega_2$. Both figures have been obtained 
by generating $10000$ stochastic trajectories by means of the Gillespie algorithm applied on Eqs. \eqref{FME} to \eqref{FME2}. Initially, the system is supposed
to be at thermal equilibrium (see text). The frequency of the normal modes as well as the coupling constants related to the system-bath interactions are the same as those
used Figure 1. The rate of energy supply is $s_k = 5$ (time arbitrary unit)$^{-1}$ for all $k$ (for (a) and (b)).}
 \end{figure}

\begin{figure}
    \includegraphics[width=4.6in]{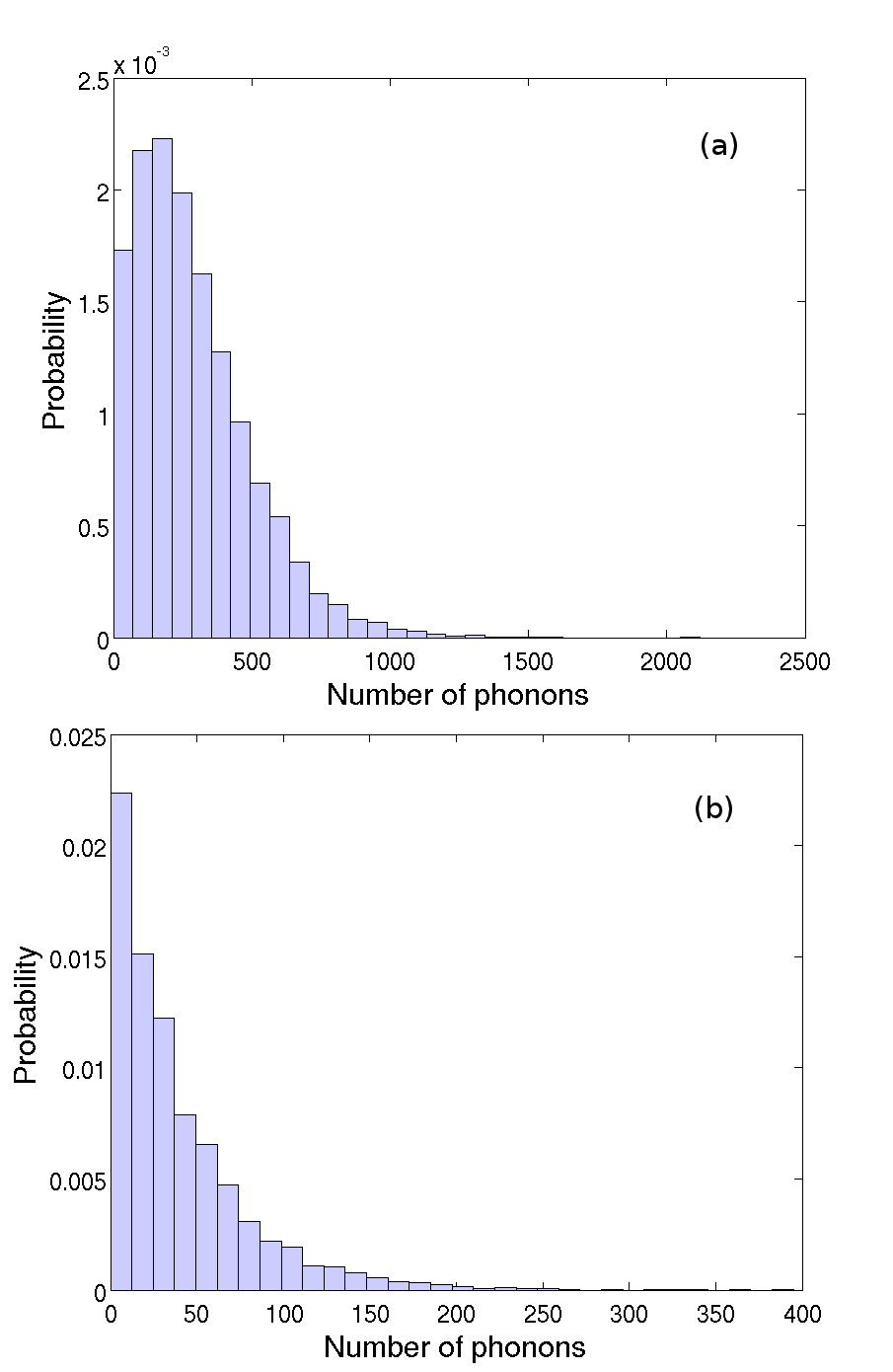}
\caption{(Color online) Same as Figure 2 with a rate of energy supply $s_k = 20$ (time arbitrary unit)$^{-1}$ for all $k$ (for (a) and (b)).}
 \end{figure}

\begin{figure}
    \includegraphics[width=4.6in]{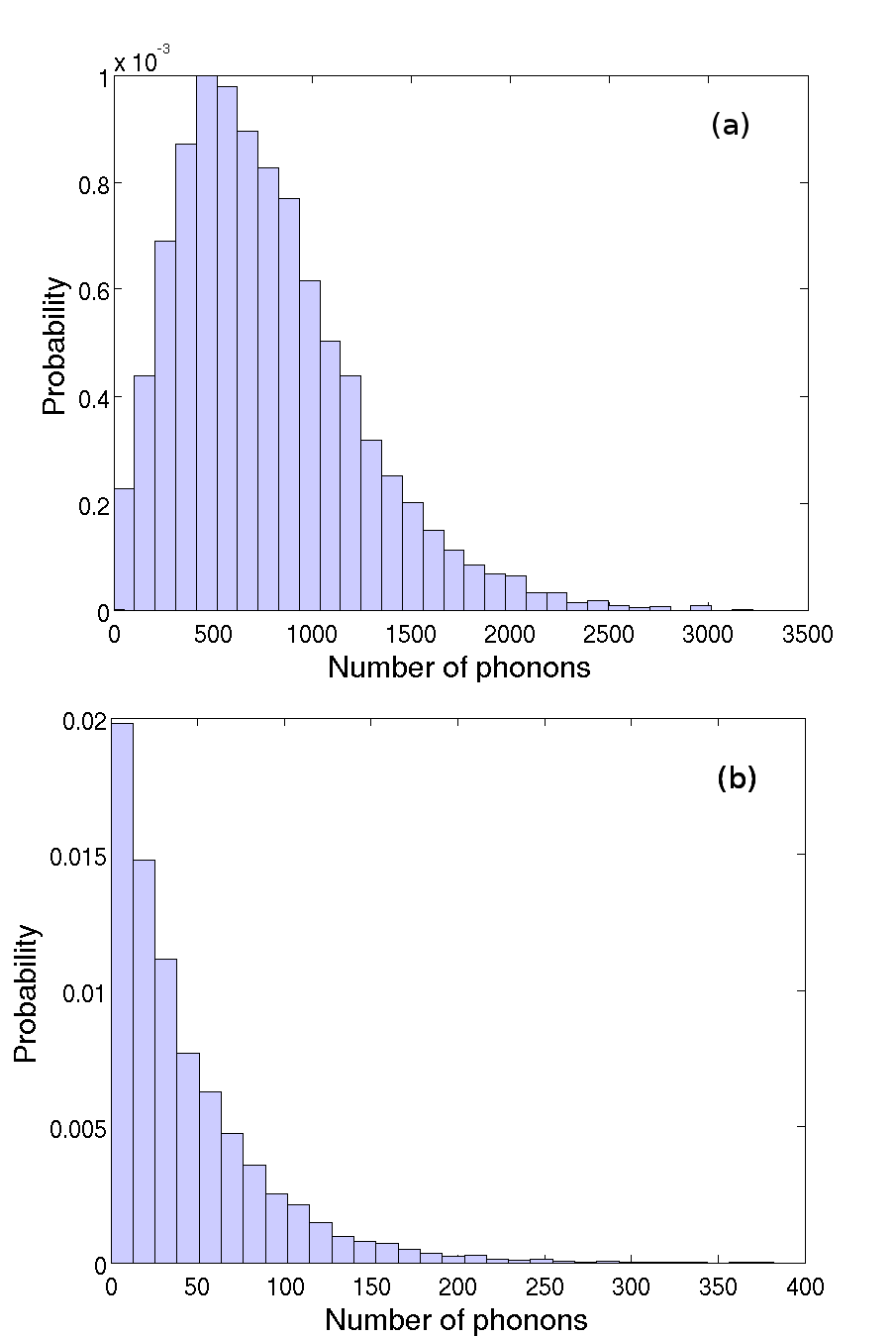}
\caption{(Color online) Same as Figure 2 with a rate of energy supply $s_k = 50$ (time arbitrary unit)$^{-1}$ for all $k$ (for (a) and (b)).}
 \end{figure}

Representative results of Fröhlich condensation using the second approach are depicted in Figure 2, 3 and 4. Each figure (histogram) was obtained by generating 
$10000$ stochastic trajectories and by storing the values of the number of phonons of a particular mode at a time $t$ when the 
system is considered to have reached the stationary state. In that regard, $t$ has been approximated beforehand using the first approach.
Again, the system is supposed to be initially at thermal equilibrium which means that each initial
condition is chosen at random following a Bose-Einstein distribution for all the modes

$$
p(n_k,t=0)= \frac{1}{Z_k} e^{-\beta\hbar \omega_k n_k}, \ \  \text{with} \ \  Z_k = \frac{1}{1 - e^{-\beta \hbar \omega_k}},
$$

for $k=1,...,z$ (rejection sampling). From a general point of view, the averages of the number of phonons deduced from the stochastic trajectories are in excellent 
agreement with those stemming from the integration of the rate equations \eqref{FRE2}. 
To this respect, let us recall that the rate equations \eqref{FRE2} can be deduced from the Fröhlich master equation
only if the correlations between the normal modes can be neglected $p({\boldsymbol{n}},t) \simeq \prod \limits_{i} p(n_i,t)$ (see section
\ref{MVPSS}). So far, with our choice of parameters, this assumption is thus confirmed independently of whether Fröhlich condensation is reached or not.

Now, we see from Figure 2, 3 and 4 how Fröhlich condensation appears statistically : for weak values of $s_k$ (Figure 2) the distributions of the number
of phonons in the first two modes of lowest frequency (four modes overall, see Figures) remain close to the Bose-Einstein distributions. Progressively,
by increasing $s_k$, we see in the first mode that not only is the average number of phonons increasing as expected but also does the variance. As a general
comment concerning the numerical results displayed here, let us remark that there is an analogy with the change of the statistical distributions of photon
occupation numbers below and above the laser transition. In fact, the photon statistics below the laser transition is a thermal equilibrium one peaked at zero, 
whereas, above the transition, the photon statistics gives a non-zero most probable value of the number of photons in the lasing mode with a well-known Poisson 
distribution. Loosely speaking something similar occurs in correspondence with the condensation transition here explored. Here, we are not interested in
deepening this analogy beyond the simple remark that the Fröhlich condensation transition displays a phenomenology typical of non-equilibrium phase
transitions in open systems.

\section{Conclusions}

In the present work, the phenomenon of Fröhlich condensation has been addressed in a semi-classical framework. With ``semi-classical'' it is meant
that the evolution of the system is described by means of classical equations with the addition of energy quantization. In the framework of 
open systems, this leads to the classical master equation replacing the evolution equation for the quantum density matrix. The quantum approaches 
hitherto proposed are based on the microscopic Hamiltonian originally proposed by Wu and Austin or on slightly modified versions of it. Here, we have proposed a 
different approach
to highlight what are the necessary hypotheses that are required to yield the Fröhlich condensation. This is not an academic 
exercise because Fröhlich phenomenon could be relevant in very different physical contexts, and, in particular, to biologically
relevant systems undergoing self-organization at different scales (from single macromolecules up to collections of cells). Thus,
the approach proposed here on the one hand has the merit of a critical reviewing of the physics underlying Fröhlich condensation and on
the other hand has the advantage of providing the time evolution of the probability density function which allows the computation of higher
moments of the number of phonons beside their average quantities. Actually, though for a system with a small number of normal modes, we 
have shown a clearcut change of the histograms of the number of phonons in correspondence with the condensation transition. Though in a somewhat loose way,
an analogy with the change of the distribution of photons below and above the laser transition has been noted.

\section*{Acknowledgments}

I warmly thank my colleagues and friends Marco Pettini and Ricardo Lima for enlightening comments and discussions. 

\appendix

\section{Detailed Balance and \textit{birth and death} process}

We consider the homogeneous Markov process $N_k(t)$ of section \ref{OVPSS}. According to the \textit{birth and death}
nature of $N_k(t)$, the stationary solution $p^s(n_k)$ of equation  \eqref{OVME} is given $\forall \ n_k \ge 0$ by
\begin{equation}\label{OVBDME}
 0 = J(n_k) -J(n_k -1),
\end{equation}

where $ J(n_k)  =   W(n_k|n_k+1) p^s(n_k+1) - W(n_k+1|n_k) p^s(n_k) $. We now sum \eqref{OVBDME} so
\begin{equation}
 0 = \sum \limits_{n_i=0}^{n_k} [J(n_i) -J(n_i -1)] =  J(n_k) - J(-1),
\end{equation}

According conditions \eqref{badcond}, $J(-1)=0$, so that the detailed balance condition
is obtained
\begin{equation}
 0 =  J(n_k) = W(n_k|n_k+1) p^s(n_k+1) - W(n_k+1|n_k) p^s(n_k), \ \forall \ n_k \ge 0
\end{equation}


\begin{thebibliography}{1}
\bibitem{note} In that regard, we refer the reader to the studies performed by Pokorný and Wu \cite{pokornybook}.
\bibitem{note2} The range $0.1-1$ THz was given by Fröhlich \cite{frohlich80} as the range of frequencies of typical vibrational biomolecular structures.
\end{thebibliography}
\end{document}